\documentclass[ ,final 
  ] {aipproc}

\layoutstyle{8x11single}
\def\Mm{M_{\rm m}}

\begin{document}

\title{Detecting the dark matter via the proper motion of $\gamma -$rays 
from microhalos}


\author{Savvas M. Koushiappas}{
  address={Theoretical Division \& ISR Division, Los Alamos National 
Laboratory, Los Alamos, NM 87545, USA}
}



\begin{abstract}
I discuss the prospects for detecting the dark matter via the proper
motion of sub--solar mass  dark matter halos in the vicinity of the solar
neighbourhood. Microhalos that survive tidal disruption  could exhibit
proper motion of order few arcminutes per year. For dark matter particles that 
couple to photons, such as the lightest supersymmetric or Kaluza-Klein particles, 
microhalos could be detected via their $\gamma$-ray photon emission from annihilations. 
A detection of proper motion of a microhalo in the $\gamma$-ray part of the spectrum 
contains not only information about the particle physics properties of the dark 
matter particle, but also provides an insight into hierarchical structure formation 
at very early times.



\end{abstract}

\maketitle


\section{Introduction}

Even though the presence of dark matter has been firmly established in
recent years, the nature of the  dark matter particle remains an
outstanding problem in particle physics. The idea that the dark matter
particle may perhaps be detected indirectly via the detection of
potential annihilation products is promising.  One particularly
interesting aspect of indirect demodulation is the detection of
$\gamma$-rays from  dark matter annihilations because it will be
readily tested with the {\it Gamma-ray Large Area Telescope}  (GLAST)
in the next few years.


The presence of sub--solar mass subhalos (microhalos) in the Milky Way
can be probed by searching for the   proper motion of $\gamma$-rays
emitted from dark matter annihilations \cite{k06,m05}. Such a detection is 
of profound
importance. First and foremost, it will be a detection of dark
matter. In addition, it will provide  information on the survival rate
of microhalos in the solar neighborhood, and will place bounds on the
mass of the dark matter particle, its annihilation cross section and
its kinetic decoupling temperature \cite{k06}.

\section{Proper motion of microhalos and expected photon flux}

In the Cold Dark Matter (CDM) paradigm, the first bound structures
form at high  redshifts. The minimum mass is set by the rms dark
matter particle velocities  dictated at kinetic decoupling at a
temperature $T_d$ \cite{schmid,h01,c01,bere03,ghs04,green05}. 
For the particular case of supersymmetric (SUSY)
dark matter, this scale is $M_{\rm min}\approx 10^{-4} ( T_d / 10 {\rm
MeV} )$ \cite{lz05}. The abundance of  microhalos in the solar neighborhood is
still under debate \cite{diemand1,diemand2,pieri,zhao,bere06,green06,tobias}. 
As such, it can be paremetrized  by assuming that
a certain fraction of the local dark matter density  ($\rho\approx
10^{-2} M_\odot {\rm pc}^{-3}$) is in microhalos in a logarithmic mass
interval \cite{k06},
\begin{equation} 
\frac{dN}{d \ln M_m d D } \approx  2 \, 
\left( \frac{ \xi}{0.002} \right) \, 
 \left( \frac{10^{-6} M_\odot}{M_m} \right) 
\left( \frac{ D }{ 0.1 {\rm pc}} \right)^2 
\end{equation}
In this parameterization, $\xi$ takes the maximum value of $\xi = 1 $
when  all of the local  dark matter density is in objects in the
logarithmic interval $M_m$. We can get an estimate on the range  of
values it may take by considering the subhalo mass function of dark
matter halos.  The subhalo mass function  has been  studied in
numerical simulations and found to be described by $dN/d \ln M \sim
M^{-1}$, normalized in a way  such that for a Milky Way size halo,
10\% of the mass of the halo is in subhalos of mass greater than
$10^7 M_\odot$. Preliminary results from N-body simulations \cite{diemand1,diemand2}, 
as well
as approximate analytical arguments  find that this mass function is
preserved down to microhalo scales, with the exception that on
sub--solar mass  scales the survival probability is reduced to only
$[10-20]$\% due to early rapid merging processes as well   as
potential interactions with stars \cite{bere06,green06,tobias}. 
In this case, the value of $\xi$ as
defined above reduces to  $\xi \approx 0.002$.

A microhalo at a distance $D$ and with a velocity ${\bf v}$ will
exhibit a proper motion of magnitude  $\mu = \tan ^{-1} [ \tau v \sin
\theta / D ] $ in a time interval $\tau$. Given a velocity
distribution  function $f(v)$ (assumed to be described by a
Maxwell-Boltzmann, see \cite{k06}), the number of microhalos  with a
proper motion greater than some threshold $\mu_{\rm th}$ is then given
by
\begin{equation} 
\frac{dN(\ge \mu_{\rm th})}{d \ln \Mm} = 4 \pi \int_0^{{D_{\rm
max}(\Mm)}}  \int_0^\infty  \frac{dN}{d \ln \Mm dD} \, f(v) \, \cos[
\theta_{\rm min}(\mu_{\rm th}, v, D)] \, dv \, dD.
\end{equation}
Here, $\theta_{\rm min}(\mu_{\rm th}, v, D) = \sin^{-1}[ D \tan
\mu_{\rm th} / v \tau ]$, and  $D_{\rm max}$ is the maximum distance
at which a microhalo is detectable (see next section).   The expected
flux of $\gamma$-rays from a microhalo of mass $M_m$ at a distance $D$
depends on the  annihilation cross section $\langle \sigma v \rangle$
and mass $M_\chi$ of the dark matter particle,  the density profile of
the microhalo, as well as the redshift at which the microhalo
collapsed, $z_{\rm form}$. Assuming an Navarro-Frenk-White profile
\cite{nfw}  and a spectrum given in \cite{b98} the  number of photons
per unit area and time with energy greater than $3 \, {\rm GeV}$ can
be written as:
\begin{equation} 
\frac{dN_\gamma(> 3 {\rm GeV})}{dA dt} \approx 4 \times 10^{-10} \, 
{\rm cm}^{-2} {\rm s}^{-1} \,  
\left( \frac{ \Mm}{10^{-6} M_\odot} \right) 
\left( \frac{0.05 {\rm pc}}{{D}} \right)^2 
\left( \frac{ \langle \sigma v \rangle }{10^{-26} {\rm cm}^3 
\, {\rm s}^{-1} } \right)
\left( \frac{ 40 {\rm GeV}}{M_\chi} \right)^2
\left( \frac{ 1 + z_{\rm form} }{1 + 70 } \right)^3
\end{equation}

\section{Detecting proper motion of $\gamma$-rays with GLAST}

\begin{figure}
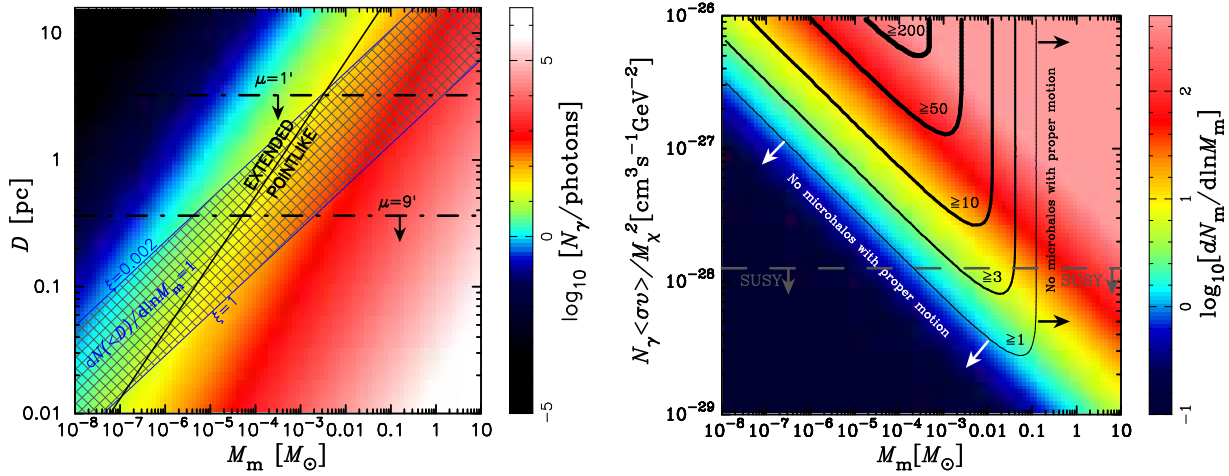

\begin{tabular}{cc}
  \includegraphics[height=.278\textheight]{figure1.ps} &
  \includegraphics[height=.28\textheight]{figure2.ps}
  \end{tabular}
  \caption{{\it Left}: Number counts of photons expected from
microhalos at different distances. The  shaded region shows the range
in distance needed to be probed in order to enclose at least 1
microhalo of mass $\Mm$. The lower (upper)  limit comes from assuming
that 100\% (0.2\%) of the local dark matter  density is in microhalos
with mass in a logarithmic interval of $\Mm$. The dot-dashed lines
show  the maximum distance at which the maximum proper motion that
could be exhibited by microhalos  is 1 \& 9 arcminutes, while the
solid line depicts the threshold over which microhalos could  be
detected as extended instead of point-like sources. The assumed
particle physics parameters  correspond to the best case scenario for
supersymmetric dark matter (see text).   {\it Right}: The number of
detectable microhalos per logarithmic mass interval as a function of  microhalo
mass and generic properties of the dark matter particle (see text).
Solid lines show iso-number contours of microhalos which will exhibit
proper motion in 2 years. Here it is assumed that 0.2\% of the  local
dark matter density is in microhalos with mass in a logarithmic
interval of $\Mm$.  The dashed line depicts the best case scenario for
supersymmetric dark matter.  }
\end{figure}

In order to detect the proper motion of microhalos that may be present
in the solar neighborhood, it is necessary  to survey the whole sky in
the $\gamma$-ray part of the spectrum. This is where GLAST is ideal in
performing  a search for the proper motion of nearby microhalos.
GLAST will survey the whole sky with an integral sensitivity of  $\sim
2 \times 10^{-10} {\rm cm}^{-2} {\rm s}^{-1}$ for photons with energy
above a few GeV.  Assuming an effective area of $2 \times 10^3 {\rm
cm}^2$, and an exposure time of 10 years (assuming the sky  will be
exposed uniformly for $5.3 \times 10^6$ seconds per year based on the
orbit configuration of  the spacecraft \cite{ritz}), the sensitivity threshold for
GLAST is $\sim 20$ photons.

The left panel of Fig. 1 shows the number of photons expected from
microhalos as a function of distance for  the same effective area and
exposure time as mentioned above. For this particular example, the mass of the dark
matter particle  was assumed to be $M_\chi = 40 {\rm GeV}$, with an
annihilation cross-section of  $\langle \sigma v \rangle = 10^{-26}
{\rm cm}^3 {\rm s}^{-1}$. This particular choice of parameters
represents  an optimistic scenario for supersymmetric dark matter.
The shaded area in Fig. 1 shows the radius of the volume needed in
order to enclose at least 1  microhalo for different values of the
parameter $\xi$. The lower limit comes from assuming that $\xi=1$,
while the upper bound corresponds to $\xi=0.002$. As can be gleaned
from Fig. 1, within reasonable distance  (e.g. within the shaded
area),  the number of photons on Earth in a 10 year exposure is
between 100-1000 for microhalos with masses  $M_m \ge 10^{-5}
M_\odot$. This number is well above the  sensitivity threshold for
GLAST,  suggesting that GLAST will be able to perform a search for
microhalos in the  solar neighborhood. If the fraction of the local
dark matter density  in microhalos of mass  less than $10^{-2}
M_\odot$ is more than 0.2\%, then there should be at least one
microhalo potentially  detectable with GLAST at a high signal-to-noise
ratio. The dot-dashed lines show the distance where the maximum
proper motion is greater or equal to 1 \& 9 arcminutes.  
The single photon angular resolution of GLAST is better
than  $\mu_0 \approx 9$ arcminutes for photons of energy above few
GeV. However,the localization of a source which  emits $N_\gamma$
photons is improved by a factor of $1/\sqrt{N_\gamma}$, thus placing a
proper motion threshold of  detection of $\mu_{\rm th} = \mu_0 /
\sqrt{N_\gamma}$. The solid line shows the threshold at which
microhalos will be resolved as extended objects in GLAST (assuming a
PSF of 9 arcminutes). In this case, the prospects  for proper motion
detection are much better because the excess photon flux from adjacent
resolution bins in  extended objects could be used to better localize
the position of the source \cite{beacon}.

The right panel of Fig. 1 shows the number
of microhalos per logarithmic mass interval that will be detectable based
on the integral sensitivity of GLAST as a function of microhalo mass
and the particle physics properties of the  dark matter particle. Here, $\xi = 0.002$, and  
the quantity $N_\gamma \langle \sigma v \rangle
/ M_\chi^2$ is  left as a free parameter (where $N_\gamma = \int_{3{\rm
GeV}}^{{M_\chi}} [dN_\gamma / dE] dE $). The particular case  of an
optimistic choice of parameters for supersymmetric dark matter is
shown with the dashed line. However, it should be emphasized that the quantity
$N_\gamma \langle \sigma v \rangle / M_\chi^2$ can attain higher
values than the optimistic case of  supersymmetry, e.g. Kaluza--Klein dark matter
\cite{ks84,bertone}. The solid lines represent iso-number contours of
the  number of microhalos detected that will also exhibit proper
motion greater than $\mu_{\rm th}$. Such a detection  can place
constraints on the properties of the dark matter particle. More
specifically, the detection of at least 1  microhalo with proper
motion places a bound on the mass of the lightest supersymmetric
particle to be less than 600  GeV, and its kinetic decoupling
temperature to be $T_d = [1-100] {\rm MeV}$ (see \cite{profumo} for more
details on the range of values of $T_d$ for supersymmetric dark matter).

The potential detection of microhalos in the solar neighborhood has
profound consequences in our understanding  of the nature of the dark
matter particle.  It is therefore essential that GLAST includes a
search for proper motion of $\gamma$-rays in the data analysis.  Such
an analysis can benefit significantly if the lifetime of GLAST is
maximized, thus increasing the baseline for  proper motion detection.


\end{document}